\documentclass[12pt]{aastex}

\usepackage{graphicx}
\usepackage{rotating}
\shorttitle{}
\shortauthors{Nesvorn\'y et al.}

\begin{document}
\baselineskip 19.pt
\title{Formation of Lunar Basins from Impacts of Leftover Planetesimals}
\author{David Nesvorn\'y$^{1,*}$, Fernando V. Roig$^2$, David Vokrouhlick\'y$^3$, William F. Bottke$^1$, \\ 
Simone Marchi$^1$, Alessandro Morbidelli$^4$, Rogerio Deienno$^1$} 
\affil{(1) Department of Space Studies, 
Southwest Research Institute, 1050 Walnut St., Suite 300,  Boulder, CO 80302, United States}
\affil{(2) Observat\'orio Nacional, Rua Gal. Jose Cristino 77, Rio de Janeiro, RJ 20921-400, Brazil}
\affil{(3) Institute of Astronomy, Charles University, V Hole\v{s}ovi\v{c}k\'ach 2, CZ–18000 Prague
  8, Czech Republic}
\affil{(4) Laboratoire Lagrange, UMR7293, Universit\'e C\^ote d'Azur, CNRS, Observatoire de la 
C\^ote d'Azur, Bouldervard de l'Observatoire, 06304, Nice Cedex 4, France}
\affil{*e-mail:davidn@boulder.swri.edu}

\begin{abstract}
The Moon holds important clues to the early evolution of the Solar System. Some 50 impact basins 
(crater diameter $D>300$ km) have been recognized on the lunar surface, implying that the early impact
flux was much higher than it is now. The basin-forming impactors were suspected to be asteroids released 
from an inner extension of the main belt (1.8--2.0 au). Here we show that most impactors were instead 
rocky planetesimals left behind at $\sim 0.5$--$1.5$ au after the terrestrial planet accretion. The 
number of basins expected from impacts of leftover planetesimals largely exceeds the number of known 
lunar basins, suggesting that the first $\sim200$ Myr of impacts is not recorded on the lunar surface. 
The Imbrium basin formation (age $\simeq 3.92$~Gyr; impactor diameter $d \gtrsim 100$ km) occurs with a 
15--35\% probability in our model. Imbrium must have formed unusually late to have only two smaller 
basins (Orientale and Schr\"odinger) forming afterwards. The model predicts $\simeq 20$ $d>10$-km impacts 
on the Earth 2.5--3.5 Gyr ago (Ga), which is comparable to the number of known spherule beds in the 
late Archean.
\end{abstract}

\section{Introduction}
In the standard model of terrestrial planet formation (Wetherill 1990), accretional collisions between 1 to 1000 km 
planetesimals lead to gradual build up of lunar- to Mars-size protoplanets that gravitationally interact and further 
grow during the late stage of giant impacts (Chambers \& Wetherill 1998). Hafnium--tungsten (Hf--W) isotopic system 
analyses indicate that the Moon-forming impact on proto-Earth happened $t = 30$--150~Myr after the appearance of the first 
Solar System solids $T \simeq 4.57$ Ga (Kleine \& Walker 2017). The new-born Moon was molten, gradually cooled down, 
and was eventually able to support impact structures on its surface. This is time zero for the lunar crater record. 
The lunar surface recorded $\simeq 50$ basin-scale impacts of $d > 20$ km bodies (Miljkovi\'c et al. 2016) since time 
zero, at least some of which formed relatively late. The radiometric dating of Apollo impact melts indicates that 
the Imbrium basin, which dominates the lunar near side, formed at $t \simeq 650$ Myr (Zhang et al. 2019). From all 
other lunar basins only Orientale and Schr\"odinger have lower accumulated density of superposed craters than 
Imbrium (Fasset et al. 2012), and must therefore be younger.

Having (at least) three basin-forming impacts happening at $t \gtrsim 650$ Myr is unexpected from the planet formation 
perspective because in the inner Solar System, where the accretion processes have relatively short timescales ($<100$ Myr), 
the impact flux should have rapidly declined over time. The ubiquity of $\sim 3.9$ Gyr ages in the Apollo samples 
has therefore motivated the impact spike hypothesis where it was assumed that the Imbrium-era impacts mark an epoch 
of enhanced bombardment (Tera et al. 1974), and prompted search for possible causes. For example, it has been suggested 
that a dynamical instability in the outer Solar System -- if it happened suitably late  -- could have destabilized asteroid 
and comet reservoirs, and produced an Imbrium-era spike (Gomes et al. 2005). The late-instability model, however, has 
fundamental problems with the survival of the terrestrial planets (Kaib \& Chambers 2016) (e.g., the fully-formed Earth 
often collides with Venus) and the delay itself (Nesvorn\'y et al. 2018), leaving the problem of the origin of lunar 
basins unresolved.   

To settle this matter, we construct a physical model for three key populations of impactors in the inner Solar System: (i) 
leftover planetesimals in the terrestrial planet zone (0.3--1.75 au), (ii) main-belt asteroids (1.75--4 au), and (iii) 
comets. The model is based on $N$-body simulations of planets and small bodies over the age of the Solar System (Sect. 2).
The simulations follow the growth of terrestrial planets, collisional and dynamical decline of populations, and record 
impacts of (i)--(iii) on the terrestrial worlds. For consistency, they include the outer planet instability 
(Tsiganis et al. 2005) -- a brief period in the early Solar System when the giant planets were scattered to their current 
orbits. The instability is assumed to happen within $t \sim 10$ Myr after the protoplanetary gas disk dispersal (Clement 
et al. 2018). We note that nature and timing of the instability are inconsequential for the main results presented here. 

\section{Methods}
\subsection{Leftover Planetesimals} 
We adopt the terrestrial planet accretion model from Nesvorn\'y et al. (2021), where protoplanets started in a 
narrow annulus ($r=0.7$--1 au) and planetesimals in a wider belt ($r \simeq 0.5$--1.5 au). The model accounted for the 
planetesimal-driven migration and dynamical instability of the outer planets (Tsiganis et al. 2005). In the 
simulation highlighted here, the instability happened at $\simeq 5.8$ Myr and the Moon--forming impact happened 
at $\simeq 40$ Myr after the gas disk dispersal, when two roughly equal-mass protoplanets -- each with mass 
$\simeq 0.5$ $M_{\rm Earth}$ -- collided. The timing of the Moon--forming impact satisfies constraints from Hf--W 
isotope systematics (Kleine \& Walker 2017). The low speed collision between two nearly equal-mass protoplanets 
falls into the preferred regime of Moon-forming impacts (Canup et al. 2022). The masses and orbits of the terrestrial 
planets were accurately reproduced in the simulations (Nesvorn\'y et al. 2021).

To determine the impact flux of leftover planetesimals on the terrestrial worlds we recorded the orbits of 
planets and planetesimals in the original simulations shortly after the Moon-forming impact. The planetesimals 
in the asteroid belt region were ignored (see below for asteroids). Mercury was not included to speed up the 
integrations; all other planets, Venus to Neptune were accounted for. We assumed that the terrestrial worlds 
were fully grown after the Moon-forming impact and their masses did not subsequently change. To increase the model 
statistics, each planetesimal was cloned by slightly altering the velocity vector ($< 10^{-6}$ fractional 
change). In total, we had nearly 130,000 planetesimal clones. The $N$-body integrator \texttt{swift\_rmvs4} 
(Levison \& Duncan 1994) was used to follow the system of planets and planetesimals over 1 Gyr. All impacts of 
planetesimals on planets were recorded by the integrator. The lunar impacts -- the Moon was not included in the 
simulations -- were obtained by rescaling the results from Earth's impacts (the gravitational focusing 
factors were computed from velocities recorded by \texttt{swift\_rmvs4}). 

We used the \texttt{Boulder} code (Morbidelli et al. 2009) to model the collisional evolution of planetesimals. 
\texttt{Boulder} employs a statistical particle-in-the-box algorithm  that is capable of simulating collisional 
fragmentation of planetesimal populations. The main input parameters of the \texttt{Boulder} code are: the (i) 
initial size distribution of the simulated populations, (ii) intrinsic collision probability $p_{\rm i}$, and (iii) 
mean impact speed $v_{\rm i}$. For (i), we adopted a broken power-law $N(>\!d) \propto d^{-\gamma}$ with $\gamma=1.5$ 
for $d < d^*$ and $\gamma=5$ for $d > d^*$, and $d^* \sim 100$ km (Youdin \& Goodman 2005, Morbidelli et al. 2009). 
We explored a wide range of initial masses ($0.001<M_0<3$ $M_{\rm Earth}$). The probabilities $p_{\rm i}(t)$ and 
velocities $v_{\rm i}(t)$ of mutual collisions between planetesimals, both as a function of time, were computed 
from the terrestrial planet simulation described above.

The \texttt{Boulder} code was run to 1 Gyr. We found that the size distribution of planetesimals rapidly changed and 
reached an equilibrium shape by only $\sim 20$ Myr. The subsequent collisional evolution was insignificant because 
the planetesimal population was reduced by a large factor. In this sense, the shape of the size distribution of 
leftover planetesimals at the time of the Moon-forming impact, and any time after that, is a fossilized imprint of 
the intense collisional grinding that happened in the first $\sim 20$ Myr (Bottke et al. 2007). The equilibrium size 
distribution shows a break at $d \simeq 100$ km, a shallow slope for $d=20$--100 km, and a slightly steeper slope 
for $d<20$ km. It is similar to that of the (scaled) asteroid belt, just as needed to explain the size distribution 
of ancient lunar craters (Strom et al. 2005). The main variability in these results is caused by the assumptions 
about (i) and the planetesimal strength $Q^*_{\rm D}$ (e.g., Benz \& Asphaug 1999). For example, we identified 
cases where the size distribution for $d=10$--100 km was slightly steeper than that of the asteroid belt. The 
steeper distribution would help to alleviate the problem with the excess of super--basins (Minton et al. 2015). 

The overall effect of collisional grinding depends on the initial mass $M_0$: stronger/weaker collisional grinding 
is found if $M_0$ was higher/lower. We characterized this dependence in detail (Nesvorn\'y et al., in preparation). 
The stronger grinding for larger initial masses leads to a situation where the population of leftover planetesimals 
for $t>20$ Myr does {\it not} sensitively depend on $M_0$. For $M_{0} > 0.1$ $M_{\rm Earth}$, we estimate 
$\sim (2.6$--$5.2) \times 10^5$ $d>10$-km planetesimals at $t=40$ Myr and use this as the standard calibration in 
Sect. 3. The effects of collisional grinding are greatly reduced for $M_{0}< 0.1$ $M_{\rm Earth}$, but the planetesimal 
population ended up to be smaller in this case (because it was already small initially). For $M_{0} < 0.03$ 
$M_{\rm Earth}$, for example, we found $<10^5$ $d>10$ km planetesimals at 40 Myr. This would imply an implausibly 
low probability of Imbrium formation ($<5$\%; Sect. 3).
\subsection{Asteroids}
A dynamical model for asteroid impactors was published in Nesvorn\'y et al. (2017a). The model used the same 
setup for the planetesimal-driven migration and instability of the outer planets as the terrestrial planet formation 
model described above. To start with, the terrestrial planets were placed on the low-eccentricity and low-inclination 
orbits. The surface density profile of asteroids was assumed to follow $\Sigma(r) \propto r^{-1}$. The radial 
profile was smoothly extended from $r>2$ au, where the model can be calibrated on observations of main belt 
asteroids (see below), to $r=1.75$--2 au. This fixed the initial number of bodies in the now largely extinct 
E-belt (Bottke et al. 2012). The results had large statistics (50,000 model asteroids) and full temporal coverage 
(4.57 Gyr). 

The flux of asteroid impactors was calibrated from today's asteroid belt. We showed that the model distribution 
at the simulated time $t=4.57$ Gyr (i.e., at the present epoch) was a good match to the orbital distribution of 
asteroids. The number of model $d>10$ km asteroids at $t=4.57$ Gyr was set to be equal to the number of $d>10$-km 
main-belt asteroids ($\simeq 8200$), as measured by the Wide-field Infrared Survey Explorer (WISE) (Mainzer et al. 2019).  
When propagated backward in time -- using the simulation results -- this provided the number of asteroids and asteroid 
impactors over the whole Solar System history.

We empirically approximated the asteroid impact flux. In terms of the cumulative number of Earth impactors with 
diameters $>d$ at times $>t$, the best fit yields  
\begin{equation}
  F(d,t) = F_1(d) \exp[ -(t/\tau)^\alpha] + F_2(d) T \ ,
  \label{flux}
\end{equation}
with  $\tau=65$ Myr, $\alpha=0.6$, $T=4570-t$ and $t$ in Myr. The first term in Eq. (\ref{flux}) accounts for the 
decline of asteroid impactors during early epochs. The second term represents the constant impact flux in the 
last 3 Gyr. There are two size-dependent factors in Eq. (\ref{flux}). $F_1(d)$ is assumed to follow the size distribution 
of main-belt asteroids; the fit to simulation results gives $F_1(10\, {\rm km}) =225$. Also, from the main belt size 
distribution, $F_1(1\, {\rm km}) = 3.0 \times 10^4$. $F_2(d)$ is calibrated on modern NEAs. Nesvorn\'y et al. (2021b) 
estimated $\sim 3$ impacts of $d>10$ km NEAs on the Earth per Gyr; we thus have $F_2(10\, {\rm km})=3\times10^{-3}$ 
Myr$^{-1}$.
\subsection{Comets}
A dynamical model for comets was developed in Nesvorn\'y et al. (2017b). To start with, a million 
cometesimals were distributed in a disk beyond Neptune, with Neptune on an initial orbit at 23 au.
The bodies were given low orbital eccentricities, low inclinations, and the surface density $\Sigma(r) 
\propto r^{-1}$. The disk was truncated at 30 au to assure that Neptune stopped migrating near 
its current orbital radius at $\simeq$30 au. The simulations were run from the time of 
the gas disk dispersal to the present epoch. The effects of outer planet (early) migration/instability, galactic 
tides, and perturbations from passing stars were accounted for in the model. The results were shown to be 
consistent with the orbital distribution of modern comets, Centaurs and the Kuiper belt.

The size distribution of outer disk cometesimals is calibrated from the number of large comets and Centaurs 
observed today, the size distributions of Jupiter Trojans and Kuiper-belt objects, and from the general condition 
that the initial setup leads to plausible migration/instability histories of the outer planets. The calibration 
gives $\sim 6 \times 10^9$ $d>10$~km and $\sim 5 \times 10^7$ $d>100$~km 
cometesimals in the original disk. The size distribution is expected to closely 
follow a power law with the cumulative index $\gamma \simeq -2.1$ for $10<d<100$ km, and have a transition 
to a much steeper slope for $d > 100$ km. The distribution is a product of the initial size distribution 
that was modified by collisional grinding. We account for the physical lifetime of comets following the method
described in Nesvorn\'y et al. (2017b).

The impact flux of comets on the terrestrial worlds is computed with the \"Opik algorithm (Bottke et al. 1994). 
The results of the \"Opik code are normalized to the initial number of comets in the original disk (see above). 
The calibrated model gives us the flux of cometary impactors over the whole history of the Solar System. An 
excellent approximation of the cumulative impact flux of comets on the Earth is
\begin{equation}
  F(d,t) = C_{\rm s}(d) \left \{  F_1 \exp [-(t/\tau_1)^{\alpha_1}] + F_2 \exp [ -(t/\tau_2)^{\alpha_2}] + 
  F_3 (4570-t) \right \}\ ,
  \label{flux2}
\end{equation}
with $F_1=F_2=6.5 \times 10^3$, $\tau_1=7$ Myr, $\alpha_1=1$, $\tau_2=13$ Myr, $\alpha_2=0.44$,
$F_3=4\times10^{-3}$ Myr$^{-1}$, $C_{\rm s}(d)=1$ for $d=10$ km, and $t$ in Myr.

\section{Results}

The integrated history of lunar impacts (Fig. 1) shows that leftover planetesimals dominated the early impact flux 
($t<1.1$ Gyr or $T>3.5$ Ga; $T$ is measured looking backward from today). Asteroids took over and produced most impacts 
in the last $\simeq 3.5$~Gyr. This may explain why the size distribution of modern lunar impactors is similar to NEAs, 
but that of early impactors was not (Strom et al. 2005, Minton et al. 2015). The cometary flux was never large enough, 
in the whole history of the inner Solar System, to be competitive. Indeed, the isotopic signatures of comets are difficult 
to find in lunar samples (Joy et al. 2012). The overwhelming majority of craters observed on the lunar surface must date back to 
$T>3.5$~Ga, when the impact flux was orders of magnitude higher than it is today. The model predicts $\simeq 100$--500 
$d>20$-km impacts for $t>30$--150 Myr ($T < 4.42$--4.54 Ga), which can be compared to only $\simeq50$ known basins
(Neumann et al. 2015). We therefore see that the number of impacts suggested by the model would be excessive if the 
Moon formed at $t=30$--150 Myr (Kleine \& Walker 2017) and recorded all large impacts since its formation. 

This suggests that the lunar record is incomplete. The Moon was fully molten when it accreted from the debris disk 
created by the giant impact on proto-Earth (Canup et al. 2022). The subsequent evolution and solidification of the Lunar 
Magma Ocean (LMO) was controlled by a number of geophysical processes, including tidal heating, formation of an 
insulating flotation crust and crust overturn (Meyer et al. 2010, Elkins-Tanton et al. 2011). 
The basins that formed when the LMO was still present would have extremely reduced topographic and crustal thickness 
signatures (Miljkovi\'c et al. 2021); they may be unidentifiable today. Suppose, for example, that the lunar surface 
started recording basin-scale impacts at $t \simeq 190$ Myr ($T \simeq 4.38$ Ga) -- the oldest known basins (e.g., 
South Pole--Aitken) would date back to this time. If so, the model implies that $\simeq 50$ $d>20$ km impacts should 
be recorded (Fig. 1), in close agreement with the number of known lunar basins. The estimated time of LMO solidification, 
$t=160$--220 Myr or $T=4.35$--4.41~Ga, is consistent with constraints from the radiogenic crustal ages and Highly 
Siderophile Elements (HSEs) in the lunar mantle (Elkins-Tanton et al. 2011, Borg et al. 2015, Morbidelli et al. 2018, 
Zhu et al. 2019). A detailed analysis of HSEs will be published elsewhere (Nesvorn\'y et al., in preparation). 

Our impact model gives general support to empirically-derived lunar chronologies (Neukum et al. 2001, Marchi 
et al. 2009, Robbins 2014), but differs in specifics (Fig. 2). 
Classically, the lunar $N_1(T)$ chronology  function -- the number of accumulated $D>1$ km craters in km$^2$ of the 
lunar surface since $T$ -- was obtained by fitting the measured crater densities on terrains with known radiometric ages. 
Neukum et al. (2001), for example, suggested $N_1(T) = a [\exp(b T) - 1] + c T$ with $a=5.44 \times 10^{-14}$ km$^{-2}$, 
$b=6.93$~Gyr$^{-1}$, and $c=8.38\times10^{-4}$ Gyr$^{-1}$ km$^{-2}$. The exponential term is identified here with the 
declining impact flux of leftover planetesimals (Fig. 1). A {\it stretched} exponential function $\exp[-(t/\tau)^\alpha]$, 
however, more accurately approximates the declining flux in our simulations. The half-life of impact decline is 
$\delta t_{\rm half} = \tau (t/\tau)^{1-\alpha} / (2 \alpha)$, where $\tau=1/b \simeq 144$ Myr and $\alpha=1$ in 
Neukum et al. (2001), and $\tau=6$ Myr and $\alpha=0.45$ in this work. This gives fixed $\delta t_{\rm half} \simeq 
72$ Myr in the classical chronology, but changing half-life in our model (e.g., $t_{\rm half}=46$ Myr for $t=200$ 
Myr and $t_{\rm half}=88$ Myr for $t=650$ Myr).\footnote{There remain significant uncertainities about the character 
of the early chronology function. For example, Robbins (2014) found a very steep decay with $t_{\rm half}=22$--30 Myr 
for $t=650$ Myr.}  

The Imbrium--era basins represent a crucial constraint on any impact chronology. In the model, $\simeq 0.31$ 
Imbrium-scale lunar impacts ($d \geq 100$ km; Miljkovi\'c et al. 2013, 2016; Schultz \& Crawford 2016)) happen for 
$t>600$~Myr. From the standard Poisson statistics, and folding in a generous $\sim 50$\% uncertainty in the model 
flux calibration (Sect. 2), we estimate that the Imbrium formation was a 15--35\% probability event. This is low, 
but not an alarmingly low probability, especially because there is a very good reason for that. There are only two 
basins, Orientale and Schr\"odinger, that formed after Imbrium. They were produced by $d \simeq 50$--64-km and 
$d \simeq 20$-km impactors, respectively (Miljkovi\'c et al. 2016, Johnson et al. 2016), which is consistent with the 
model expectation of $\sim 2$ basin-scale lunar impacts for $t>600$ Myr (Fig. 1). Having only two smaller basins 
with post-Imbrium formation ages, however, is surprising. For example, if the current asteroid--belt size distribution is 
adopted for reference, there should (on average) be $\sim 7.4$ $d>20$-km impacts for every $d>100$-km impact. This 
suggests that the Imbrium basin formed unusually late, by chance, to have only two smaller and younger basins than 
Imbrium (there would be many more younger/smaller basins otherwise), and justifies the 15--35\% probability of Imbrium 
formation quoted above. In fact, our impact chronology is (nearly) optimal to satisfy the Imbrium-era constraints 
(Fig. 3). 

The Earth receives $\simeq 20$ times more impacts than the Moon. When a large impactor strikes the Earth, it produces 
a vapor-rich ejecta plume. As the plume cools down, glassy spherules form and fall back, producing a global layer 
that can be several millimeters to many centimeters thick. Some $\sim 16$ spherule beds have been found in the 
late Archean period ($T=2.5$--3.5 Ga; Marchi et al. 2021), although preservation biases and incomplete sampling 
may be an issue. At least some of these layers may have been produced by very large, $d \sim 50$ km impactors, 
but most are thought to record $d>10$ km impacts (Marchi et al. 2010; Johnson et al. 2016b). This can be compared 
with the model predictions. We estimate $\simeq 20$ $d>10$-km impacts on the Earth for $T=2.5$--3.5 Ga (Fig. 4),
of which $\sim 2$ should be $d>50$ km. The leftover planetesimals and main-belt asteroids contribute equally to 
impacts in this time interval ($\sim 10$ impacts each). Whereas the asteroid impacts should be more uniformly 
spread over the late Archean, nearly all planetesimal impacts happen before $\sim 3$ Ga. The model gives $\sim 10$ 
and $\sim 2$ $d>10$-km asteroid impacts on the Earth in the past 2.5 and 0.6 Gyr, respectively, which is 
consistent with the current impact flux of large NEAs (Nesvorn\'y et al. 2021).

Even though comets were dwarfed by planetesimals and asteroids in terms of the overall bombardment, evidence for 
cometary impacts can be found in the composition of Earth's atmosphere (Marty et al. 2016). Comets start to be released 
from  the trans-Neptunian region near the onset of the outer planet instability. The cometary impact profile 
is found here to be more extended in time than previously thought (morbidelli et al. 2018), with 10\% of cometary 
impacts happening at $>55$~Myr and 1\% at $> 370$ Myr after the instability. Adopting a case with the outer planet 
instability at $t<10$~Myr and Moon formation at $t \sim 50$ Myr, we estimate that the Earth would have accreted 
$\sim 2 \times 10^{22}$ g of cometary material after the Moon-forming impact. This is consistent with comets being 
the main source of noble gases in the Earth atmosphere (Marty et al. 2016; but a negligible source of Earth's water -- 
the mass of Earth's oceans is $1.4 \times 10^{24}$ g). In fact, the atmospheric abundance of noble gases can be 
used to constrain the delay between the outer planet instability and Moon formation, $\Delta t$. We estimate 
$20 < \Delta t < 60$ Myr, with the exact value depending on the physical lifetime of comets (Sect. 2). This 
suggests, if the instability happened very early ($t < 10$ Myr; Clement et al. 2018, Liu et al. 2022), that the 
Moon must have formed early as well ($t < 70$ Myr or $T>4.5$ Gyr; Thiemens et al. 2019).

In a broader context, our work highlights the importance of impacts for the early Earth. The bulk of accreted mass was 
brought by 80--400 $d>100$ km (Imbrium--scale) impacts and the stochastic accretion of very large planetesimals 
($d \gtrsim 1$,000 km), as needed to explain Earth's HSEs (Bottke et al. 2010). If the surface of Earth's Hadean crust was 
widely reprocessed by these impacts, this could explain the age distribution of Hadean zircons and absence of terrestrial 
rocks older than 4.3 Gyr (Marchi et al. 2014). There is roughly a 50\% chance that the last $d>100$ km impact on the 
Earth happened as late as $T \lesssim 3.5$ Ga. In total, the early Earth received $\sim 2$,000--10,000 $d>10$ km 
impacts, each representing the magnitude of the Cretaceous--Paleogene (K/Pg) extinction-scale event (Alvarez et al. 1980). 
By $T \sim 3.7$ Ga, when we have the earliest firm evidence for biotic life (Nutman et al. 2016), the impact flux has 
already declined $\sim 1000$ times, and the mean interval between K/Pg-scale impacts stretched to $>1$ Myr. 
%Large impacts also affected the evolution of early--Earth's atmosphere (Sinclair et al. 2020), possibly causing 
%delayed and variable atmospheric oxidation (Marchi et al. 2021).    
%In the case highlighted here, the Moon-forming impact happens at $t \simeq 40$ Myr, but the results are insensitive 
%to that. 

\section{Conclusions}

The main results of this work can be summarized as follows.

\begin{itemize}
\item The leftover planetesimals produce most lunar impacts in the first 1.1 Gyr ($t<1.1$ Gyr or $T>3.5$ Ga).
Asteroids produce most impacts in the last 3.5 Gyr. The transition from leftover planetesimals to asteroids 
has been imprinted in the crater size distributions (Strom et al. 2005, Head et al. 2010, Orgel et al. 2018). 
The comet contribution to the crater record is found to be insignificant (for the early instability case 
adopted here). 
\item Some $500$ $d>20$-km planetesimals from the terrestrial planet zone (0.5--1.5 au) are expected to 
impact the Moon since its formation. The early crater record must have been erased because the lunar surface 
was unable to support basin-scale impact structures. The $\sim 50$ known lunar basins formed after $t \simeq 
160$--215 Myr ($T \lesssim 4.36$--4.41 Ga). This is consistent with the long-lived LMO (Morbidelli et al. 
2018, Zhu et al. 2019). The South Pole--Aitken basin should date back to $T = 4.36$--4.41 Ga.
\item About two lunar basins are expected to form for $t \gtrsim 650$ Myr ($T \lesssim 3.92$ Ga). The Imbrium 
basin formation ($T \simeq 3.92$~Ga, $d \gtrsim 100$ km impactor) is estimated to happen with a 15--35\% 
probability in our model. Imbrium should have formed unusually late, relative to the expectations from the lunar 
impact chronology, to have only two smaller/younger basins than Imbrium (Orientale and Schr\"odinger); there 
would be many more younger/smaller basins otherwise.  
\item The lunar chronology can be given as a sum of two terms: the stretched exponential
function (the leftover planetesimal branch) and a constant (the asteroid or NEA branch). This is similar to the 
classical (empirical) crater chronologies (Neukum et al. 2001, Marchi et al. 2009, Robbins 2014), except that 
the cratering rate profile in the first $\sim 1$ Gyr had a longer tail than the exact exponential. 
This can lead to modest, $\sim 50$ Myr differences in the estimates of lunar basin ages.
\item Our model predicts $\simeq 20$ $d>10$-km impacts on the Earth for $T=2.5$--3.5 Ga. This is similar 
to the number of known spherule beds in the late Archean (Bottke et al. 2012, Johnson et al. 2016, Marchi 
et al. 2021). Both the leftover planetesimals and main-belt asteroids contribute to impacts in this time 
interval. Whereas the asteroid impacts were more uniformly spread over the late Archean, nearly all 
planetesimal impacts should have happened before 3 Ga.
\end{itemize}

\section{A note on previous work}

Bottke et al. (2007) studied lunar impacts of leftover planetesimals and concluded that the leftover 
planetesimals \textit{cannot} produce basin--scale impacts on the Moon during the Imbrium era, because the 
collisional and dynamical decline of planetesimal impactors was presumably too quick. Comparing their 
impact profile with the ones obtained here, we identify the main reason behind this: the planetesimal 
population in Bottke et al. (2007) dynamicaly decayed by two orders of magnitude in the first 100 Myr. We do not find 
any such strong initial trend in our work. The problem in question is most likely related to the approximate 
nature of initial conditions in Bottke et al. (2007), where the present-day near-Earth asteroids (NEAs) were 
used as a proxy for terrestrial planetesimals. We note that modern NEAs have relatively short dynamical 
lifetimes. 

Brasser et al. (2020) developed a dynamical model for lunar impactors with four different components: E-belt, 
asteroid belt, comets and leftover planetesimals. The main difference relative to our work is that their model 
was {\it not} absolutely calibrated from independent means. To match the lunar impact constraints in their 
model, Brasser et al. (2020) increased $\sim10$ times the population of the E-belt (cf. Bottke et al. 2012), 
and suggested that the E-belt asteroids were the main source of basin-forming impacts. The best-fit 
contribution of planetesimals was found to be negligible; leftovers were therefore concluded to be only a minor
 source of lunar impacts. This can be compared to our work where we find the dominant role of leftover 
planetesimals.

\acknowledgements
The work of D.N. was supported by the NASA Emerging Worlds program. F.R. acknowledges support from 
the Brazilian National Council of Research - CNPq. The work of D.V. was supported by the Czech Science Foundation 
(grant number 21--11058S). A.M. 
received funding from the European Research Council (ERC) under the European Union's Horizon 2020 research and 
innovation program (grant No. 101019380 HolyEarth). R.D. acknowledges support from the NASA Emerging Worlds program, 
grant 80NSSC21K0387. The simulations were performed on NASA’s Pleiades Supercomputer. We greatly appreciate the 
support of the NASA Advanced Supercomputing Division. The {\tt SWIFT} code is available on 
\texttt{www.boulder.swri.edu/\~{}hal/swift.html}. The data that support the plots within this paper and other 
findings of this study are available from the corresponding author upon reasonable request.

\clearpage
\begin{figure}
\epsscale{1.5}
\hspace*{-4.5cm}\plotone{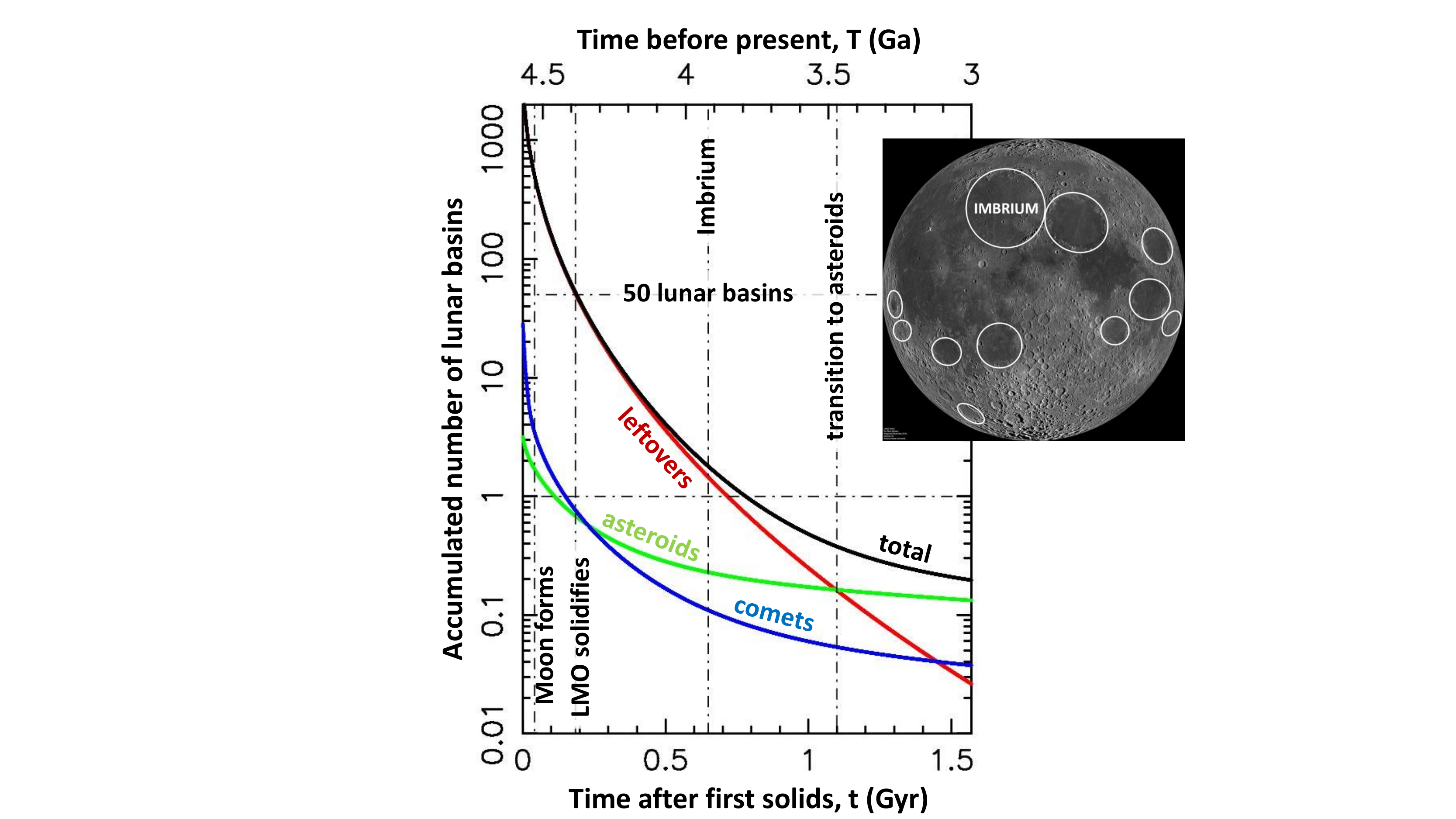}
% gr.impact_moon.f
\caption{Early impacts of diameter $d>20$-km planetesimals on the Moon match the number of known lunar 
basins -- on the assumption that the basin record started at $t \simeq 190$~Myr. The plot shows the accumulated 
number of impacts after time $t$, where $t=0$ is the birth of the Solar System, and $t \simeq 4.57$ Gyr is the 
present time. The profiles have a declining trend as younger surfaces accumulate fewer impacts. The 
planetesimal, asteroid and comet profiles are shown by red, 
green and blue lines, respectively; the black line is the total impact flux. The vertical dash-dotted lines 
show the lower bound on the Moon-forming impact ($t \sim 30$~Myr), estimated start of the lunar basin 
record ($t \simeq 190$ Myr or $T \simeq 4.38$ Ga), Imbrium formation ($t \simeq 650$ Myr or $T\simeq3.92$ Ga), 
and transition from the planetesimal-dominated to asteroid-dominated impact stages ($t \simeq 1.1$ Gyr or $T 
\simeq 3.5$ Ga). The inset shows some of the lunar near-side basins.}
\label{chrono}
\end{figure}

\clearpage
\begin{figure}
\epsscale{1.6}
\hspace*{-4.0cm}\plotone{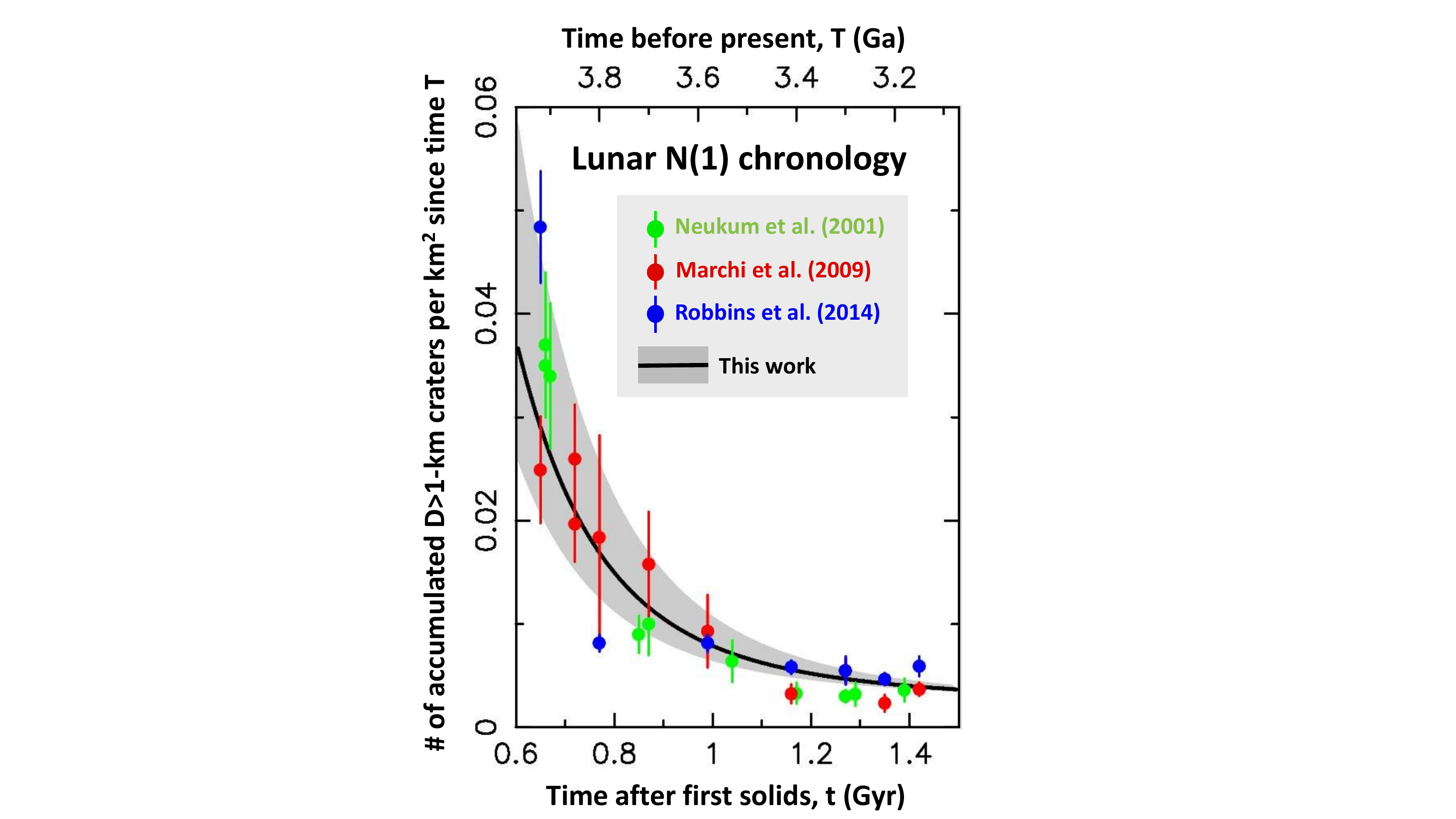}
\caption{Our model $N_1$ chronology matches crater counts on lunar terrains with known radiometric ages. The bold 
solid line is $N_1(T)=a \exp [ -(t/\tau)^\alpha] + c T$ with $t=4570-T$ ($t$ and $T$ in Myr), $a = 94$ km$^{-2}$, 
$\tau=6$ Myr, $\alpha=0.45$, and $c=10^{-6}$ km$^{-2}$ Myr$^{-1}$; the shaded area indicates the model uncertainty. 
The green (Neukum et al. 2001), red (Marchi et al. 2009), and blue dots (Robbins 2014) with associated error 
bars show different $D>1$-km crater counts. The impact model also implies $\simeq 25$ $D>20$-km craters per 
$10^6$ km$^2$ for $T=3.92$~Ga, in a close match to $N_{20}=26 \pm 5$ reported for the Fra Mauro/Imbrium 
highlands (Orgel et al. 2018). The Nectaris basin with $N_{20} \simeq 170$ is estimated here to be 
$T=4.21$--4.29 Gyr old.}
\label{lns}
\end{figure}

\clearpage
\begin{figure}
\epsscale{1.4}
\hspace*{-2.6cm}\plotone{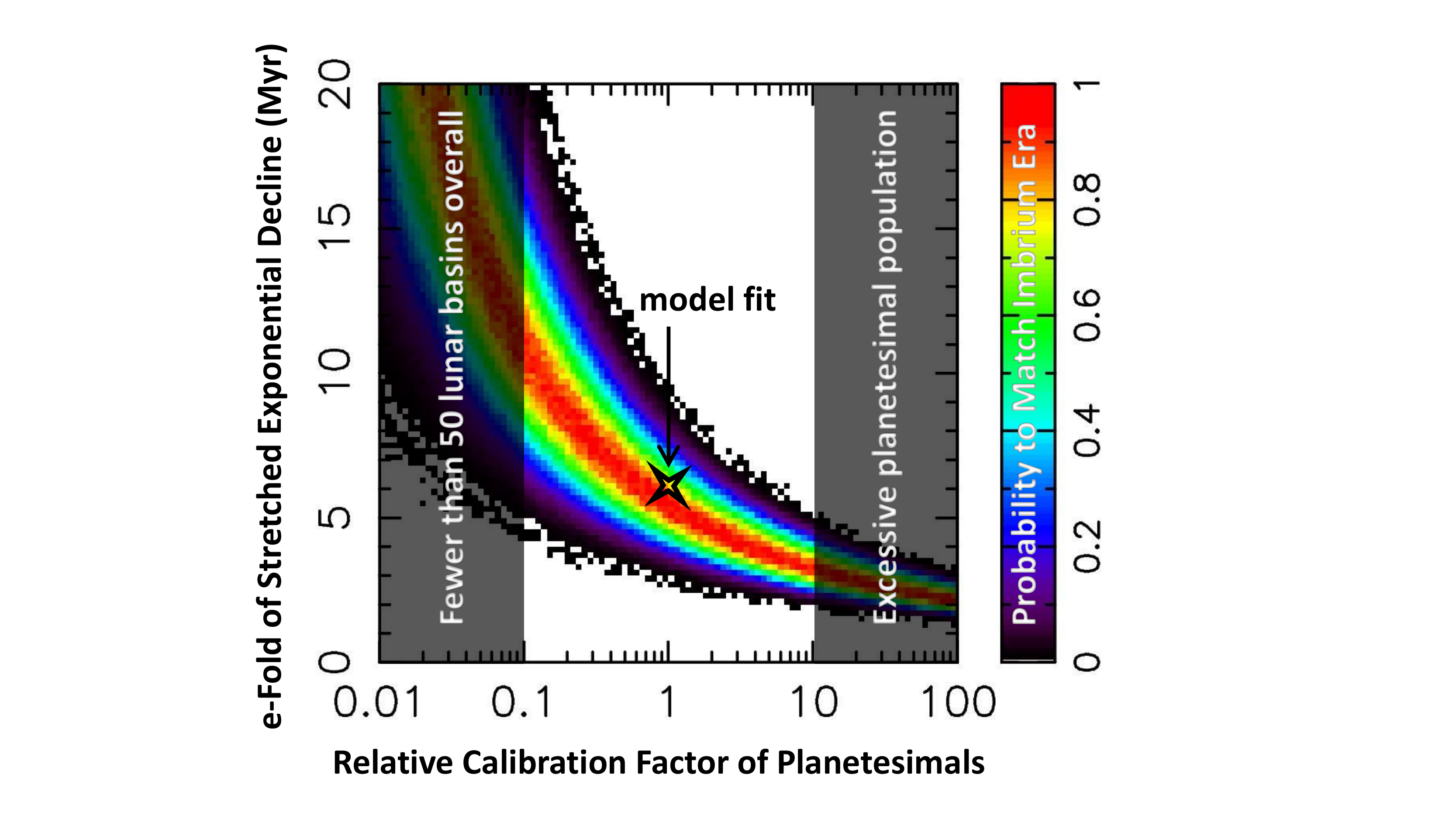}
\caption{The model impact chronology satisfies Imbrium-era constraints. The rainbow plot shows the normalized 
probability for Imbrium to form late ($600<t<900$ Myr, $d>100$~km) {\it and} have exactly two basins (Orientale 
and Schr\"odinger, $d>20$ km impactors) younger than Imbrium. We fix $\alpha=0.45$, and test 
different chronologies by varying the $e$-fold $\tau$ and relative calibration factor $C_{\rm r}$, where $C_{\rm r}=1$ 
corresponds to the standard calibration of leftover planetesimals (Sect. 2.1). We generate a statistically large number of 
random impact sequences ($\sim 10^5$) for each pair $(C_r,\tau)$ and evaluate the likelihood of satisfying 
the condition described above. The likelihood is low if $C_r$ and/or $\tau$ are small, because it is difficult
to form Imbrium. It is low if $C_r$ and/or $\tau$ are large, because more than one Imbrium forms and/or 
many smaller basins form after Imbrium. The star symbol denotes $C_{\rm r}=1$ and $\tau=6$ Myr obtained 
from a fit to our simulation of leftover planetesimals.}
\label{imbrium}
\end{figure}

\clearpage
\begin{figure}
\epsscale{1.5}
\hspace*{-3.5cm}\plotone{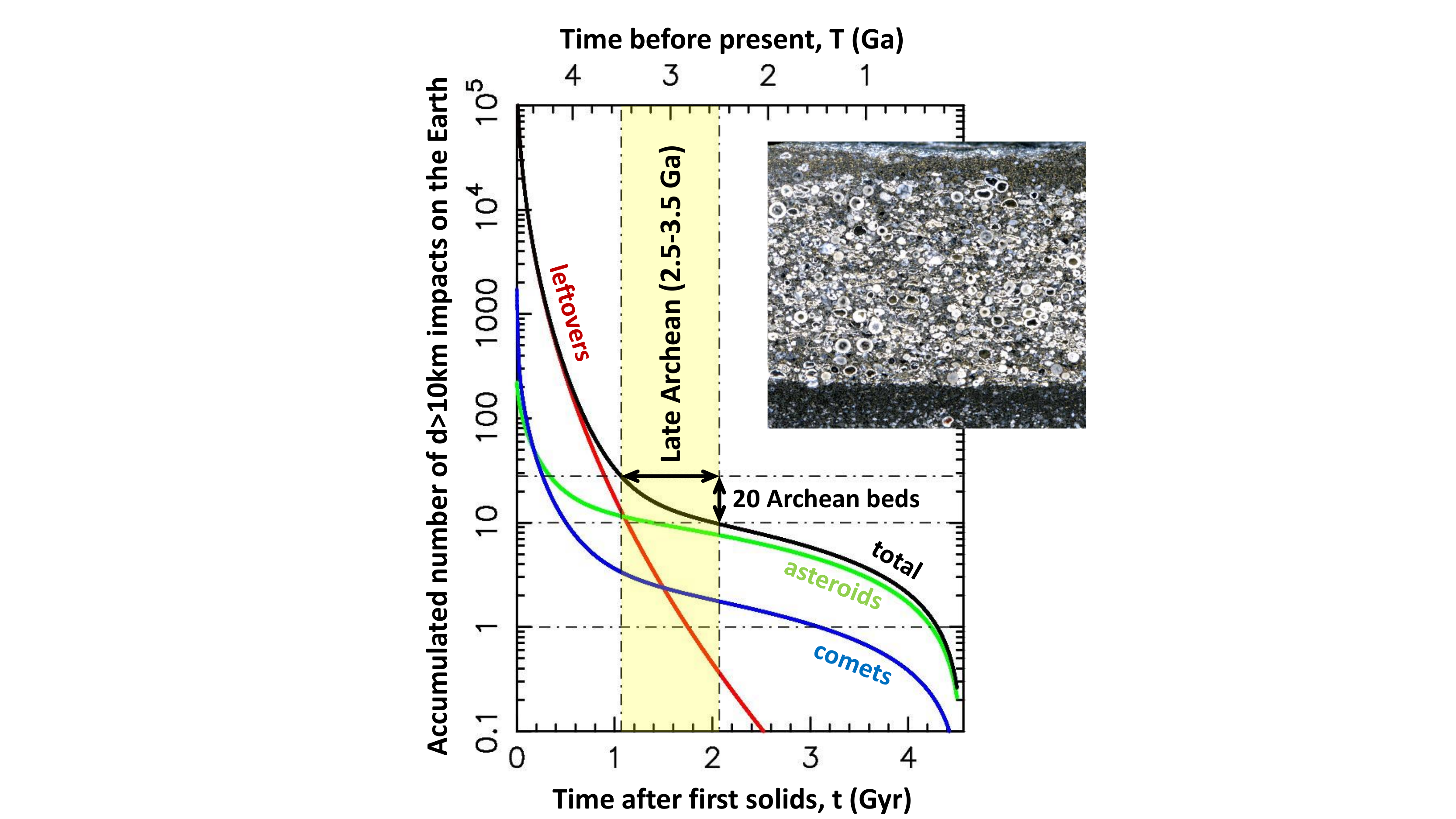}
\caption{Model impacts of diameter $d>10$ km bodies on the Earth match the number of Archean spherule 
beds. The vertical dash-dotted lines outline the late Archean period, $T = 2.5$--3.5 Ga, where the model
predicts $\simeq 20$ $d>10$-km impacts on the Earth. The sixteen known Archean spherule beds occur in two distinct 
time intervals, 2.4--2.7 Ga and 3.2--3.5 Ga (Marchi et al. 2021). If this is indicative of the average flux in the 
late Archean, there should be $\sim 14$ additional spherule beds at 2.7--3.2~Ga. If so, the number of $d>10$ km 
impacts in the model would represent $\sim 60$\% of the total number of spherule beds. For reference, some $\sim 43$ 
$d>7$ km impacts on the Earth are expected in the model at 2.5--3.5~Ga. The inset shows the Monteville spherule 
layer in South Africa (Reimold \& Koeberl 2014).} 
\label{earth}
\end{figure}

\end{document}